\documentclass[%
superscriptaddress,
bibnotes,
 amsmath,amssymb,
 aps,
 reprint,
prb,
]{revtex4-2}

\usepackage{graphicx}
\usepackage{dcolumn}
\usepackage{bm}
\usepackage{physics}
\usepackage{siunitx}
\usepackage[normalem]{ulem}
\usepackage{xprintlen}
\usepackage[T1]{fontenc}
\usepackage{romannum}

\usepackage{layouts}
\usepackage{float}
\usepackage{amsmath}
\newcommand{\RomanNumeralCaps}[1]
    {\MakeUppercase{\romannumeral #1}}

\usepackage[caption=false]{subfig}

\usepackage{todonotes}

\usepackage[colorlinks=true, citecolor=blue, linkcolor = black, urlcolor = black]{hyperref}
\usepackage[capitalize]{cleveref}   

\begin{document}

\title{Telecom quantum dots on GaAs substrates as integration-ready high performance single-photon sources}

\author{Beatrice Costa}
\email{beatrice.costa@tum.de}
\thanks{ORCID: 0009-0007-9528-6206}
\affiliation{Department of Electrical and Computer Engineering, TUM School of Computation, Information and Technology, Technical University of Munich}
\affiliation{Munich Center for Quantum Science and Technology (MCQST)}
\author{Bianca Scaparra}
\affiliation{Department of Electrical and Computer Engineering, TUM School of Computation, Information and Technology, Technical University of Munich}
\affiliation{Munich Center for Quantum Science and Technology (MCQST)}
\author{Xiao Wei}
\affiliation{Department of Electrical and Computer Engineering, TUM School of Computation, Information and Technology, Technical University of Munich}

\affiliation{Munich Center for Quantum Science and Technology (MCQST)}
\author{Hubert Riedl}
\affiliation{Walter Schottky Institute, TUM School of Natural Sciences, Technical University of Munich}
\author{Gregor Koblmüller}
\affiliation{Munich Center for Quantum Science and Technology (MCQST)}
\affiliation{Walter Schottky Institute, TUM School of Natural Sciences, Technical University of Munich}
\affiliation{Institute of Physics and Astronomy, Technical University Berlin}
\author{Eugenio Zallo}
\affiliation{Munich Center for Quantum Science and Technology (MCQST)}
\affiliation{Walter Schottky Institute, TUM School of Natural Sciences, Technical University of Munich}
\author{Jonathan Finley}
\affiliation{Munich Center for Quantum Science and Technology (MCQST)}
\affiliation{Walter Schottky Institute, TUM School of Natural Sciences, Technical University of Munich}
\author{Lukas Hanschke}
\affiliation{Department of Electrical and Computer Engineering, TUM School of Computation, Information and Technology, Technical University of Munich}
\affiliation{Munich Center for Quantum Science and Technology (MCQST)}
\author{Kai Müller}
\affiliation{Department of Electrical and Computer Engineering, TUM School of Computation, Information and Technology, Technical University of Munich}
\affiliation{Munich Center for Quantum Science and Technology (MCQST)}

\date{\today}

\begin{abstract}

The development of deterministic single photon sources emitting in the telecommunication bands is a key challenge for quantum communication and photonic quantum computing.
Here, we investigate the optical properties and single-photon emission of molecular beam epitaxy grown semiconductor quantum dots emitting in the telecom O- and C- bands. The quantum dots are embedded in a InGaAs matrix with fixed indium content grown on top of a compositionally graded InGaAs buffer. This structure allows for the future implementation of electrically contacted nanocavities to enable high-quality and bright QD emission. In detailed optical characterizations we observe linewidths as low as \SI{50}{ \micro eV}, close to the spectrometer resolution limit, low fine structure splittings close to \SI{10}{ \micro eV}, and $g^{(2)} (0)$ values as low as $0.08$. These results advance the current performance metrics for MBE-grown quantum dots on GaAs substrates emitting in the telecom bands and showcase the potential of the presented heterostructures for further integration into photonic devices.

\end{abstract}

\maketitle

\section{Introduction}

Future advancements in quantum communication require high-performance single-photon sources, capable of generating pure, indistinguishable photons with high efficiency. Semiconductor quantum dots (QDs) emitting at 950 or \SI{785}{ nm} have emerged as promising candidates for single-photon sources, due to their excellent optical properties and potential for integration into scalable photonic devices \cite{senellart2017high,  hanschke2018quantum, sbresny2022stimulated, da2021gaas, schöll2019resonance, vajner2022quantum, basso2021quantum}. Significant progress has recently been reported in the optimization of this system, such as spectral tuning mechanisms, cavity integration, and strain engineering \cite{gazzano2013bright, kolatschek2019deterministic, sapienza2015nanoscale}.\\
\indent In fiber-based quantum networks, the communication range is limited by optical attenuation, which scales exponentially with distance. This constraint makes the realization of sources emitting in the telecommunication bands crucial, as they minimize photon losses in optical fibers and enable practical long-distance quantum communication \cite{gyger2022metropolitan}. Among the most promising candidates for telecom applications are InAs QDs grown on InP substrates, as they naturally emit in the C-band \cite{benyoucef2013telecom, holewa2022droplet, muller2018quantum}.
Specifically, the performance of QDs as a single-photon source can be improved by embedding them in photonic resonators to enhance photon extraction and coherence properties \cite{holewa2024high, phillips2024purcell, jeon2022plug}. While InP-based QDs exhibit favorable emission wavelengths due to their lower lattice mismatch, further enhancing the extraction efficiency using distributed Bragg reflectors (DBRs) is challenging due to the relatively low refractive index contrast of the materials used for InP-based DBRs compared to GaAs-based ones.\\
\indent On the other hand, GaAs substrates are less brittle than InP ones and allow for the growth of lattice-matched binary DBR systems with high refractive index contrast.
Importantly, for GaAs substrates, an effective method for shifting QD emission to the telecom bands is to grow compositionally graded InGaAs layers directly beneath the InAs quantum dot layer \cite{paul2017single, wronski2021metamorphic, scaparra2023structural, wyborski2023impact}. However, plastic relaxation of the lattice constant along the graded layer leads to the formation of dislocations, thereby complicating the implementation of electrically contacted devices. Additionally, the QD layer is grown atop a residually strained region within the graded structure, further complicating device integration \cite{scaparra2023structural, sacedon}.
Despite these challenges, a nonlinear grading profile was proposed to reduce the thickness of the graded layer and thus fabricate a circular Bragg grating (CBG) to enhance the extraction efficiency. Nonetheless, the relaxed portion of the layer was included in the final resonator, hence limiting the implementation of additional electrical contacts \cite{sittig2022thin}.

A potential solution that addresses the limits associated with the direct growth of QDs on a graded layer is to embed the QDs within a relaxed InGaAs matrix grown on top of the compositionally graded layer, resulting in reduction of both dislocation density and residual strain close to the QD layer \cite{semenova2008metamorphic,SOROKIN201683TEM, scaparra2024broad}. 
Recently, such an approach has been optimized to obtain bright QD emission in the telecom O- and C-bands \cite{scaparra2024broad}.
However, comprehensive studies of the optical properties of this heterostrucrure such as optical linewidths, fine structure splittings (FSSs), lifetimes, and the single-photon nature of the emission are still missing.

\begin{figure}[!t]
\centering
\includegraphics[width=0.5\textwidth]{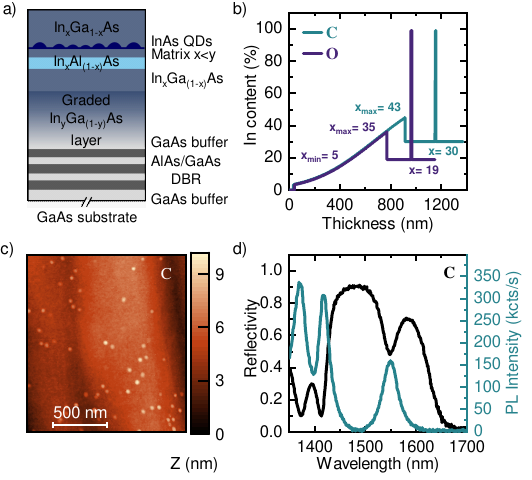}
\caption{a) Schematic of the sample structure: the InAs QDs are embedded in an InGaAs matrix with fixed indium content grown on top of a compositionally graded InGaAs layer. An AlAs/GaAs distributed Bragg reflector (DBR) is grown on a (001) GaAs substrate. b) Profile of the In content across heterostructures \textbf{O} (purple) and \textbf{C} (green) excluding the DBRs layers. c) AFM scan of a sample similar to heterostructure \textbf{C} showing uncapped QDs grown on top of an InGaAs matrix. d) Reflectivity measurements (black) and ensemble PL spectrum of the QD emission (green) measured via non-resonant excitation at \SI{10}{K} from sample \textbf{C}. Both spectra are recorded at the same position on the wafer.}
\label{figure1}
\end{figure}

In this paper, we investigate the optical properties of InAs QDs embedded in mostly relaxed InGaAs matrices grown on top of graded InGaAs buffers placed on nearly lattice-matched DBRs with emission across the telecom O- and C-bands. Specifically, micro-photoluminescence ($\mu$-PL) and time-resolved characterizations allow us to analyze key metrics such as linewidth, FSS, and excitonic lifetime, and to demonstrate the single-photon nature of the emission via photon statistics measurements.

\section{Sample structure}

The studied samples were grown with a solid source Veeco Gen \RomanNumeralCaps{2} molecular beam epitaxy (MBE) system on undoped GaAs(001) substrates.
Two different samples were grown, where the layer thicknesses and In composition were optimized for emission of the QDs in the O- and C-band, respectively. Throughout this paper, these two samples are labelled sample \textbf{O} and \textbf{C}.
Figure~\ref{figure1}a shows the layer structure of the studied samples.
First, a \SI{200}{ nm}-thick GaAs buffer layer was grown on the GaAs (001) substrate to achieve an optimal epitaxial surface. 
Subsequently, 10 alternating GaAs/AlAs DBR pairs were incorporated into each sample with their thicknesses designed to enhance the extraction efficiency of the QD emission in the center of the respective telecom band.

To ensure that the total optical thickness between the DBRs and the semiconductor–air interface forms a $3\lambda$ cavity, an additional GaAs buffer layer is deposited atop the DBRs, with thicknesses of \SI{40}{ nm} and \SI{30}{nm} for samples \textbf{O} and \textbf{C}, respectively. 

Figure~\ref{figure1}b shows the In content profiles across both samples, excluding the DBR layers. The reported In profiles were calculated from the In-flux equivalent growth rate as a function of the In cell temperature at fixed Ga flux \cite{scaparra2024broad}. 
During the deposition of the graded In$_{\text{y}}$Ga$_{\text{1-y}}$As layers, the Ga growth rate was kept at \SI{1}{\text{ \AA/}s}, while the In growth rate was increased from \SI{0.05}{\text{ \AA/}s} to \SI{0.53}{\text{ \AA/}s} for sample \textbf{O} and up to \SI{0.75}{\text{ \AA/}s} for sample \textbf{C}. This resulted in maximum In contents of $35\%$ and $43\%$, respectively. Using an In grading rate similar to the one reported in Ref.~\cite{scaparra2024broad}, the thicknesses of the compositionally graded layers were determined to be \SI{730}{ nm} and \SI{880}{ nm} for sample \textbf{O} and \textbf{C}, respectively.\\
\indent An equivalent thickness of 2.2 monolayers (MLs) of InAs forming the QDs were embedded within an InGaAs matrix with In compositions of $19\%$ ($30\%$) for sample \textbf{O} (sample \textbf{C}). The substrate rotation was stopped during the growth of the QD layer to achieve a QD density gradient. The QD layers were positioned in an antinode of the standing electric field, i.e., \SI{190}{ nm} (sample \textbf{O}) and \SI{220}{ nm} (sample \textbf{C}) from the surface.
In order to prevent further propagation of dislocations into the matrix, a \SI{20}{ nm}-thick InAlAs layer, with In composition matching that of the InGaAs matrix, were introduced in both samples. Further details regarding the growth process can be found in Refs.~\cite{scaparra2024broad, scaparra2023structural}. \\
\indent Figure~\ref{figure1}c shows an AFM topography scan of an uncapped reference sample consisting of InAs QDs on an InGaAs matrix grown on top of the graded buffer, taken from a high density (approximately $2.2 \times 10^9 \text{ cm}^{-2}$) region.
The surface undulations with a width of \SI{1}{ \micro m} and an height of \SI{10}{ nm} are typical for metamorphic structures \cite{Andrews2002crosshatch,paul2017single,scaparra2023structural} where QD nucleation is more favorable on convex parts of the surface \cite{sittig2022thin,scaparra2023structural}.\\
\indent After the growth, reflectivity measurements and ensemble PL measurements are used to confirm the correct design of the cavity. As an example, figure~\ref{figure1}d shows the normalized reflectivity (black) and PL (green) spectra from sample \textbf{C} recorded at a temperature of \SI{10}{K} at the same position on the wafer. The reflectivity corresponds to the typical reflectivity of a DBR stop band, with a cavity dip in the center. The layer structure selectively suppresses light out-coupling within specific wavelength ranges while enhancing it near the cavity dip. Notably, the PL spectrum of the QDs, presented alongside the reflectivity data, reveals a pronounced enhancement of QD emission near the cavity dip at \SI{1550}{nm}. This is due to the positioning of the QD layer near an anti-node of the standing wave field.

\section{Optical characterization}
In the following, we present a detailed characterization of the optical properties of single QDs under above band excitation, including power-dependent, polarization-resolved, and time-resolved PL measurements. From these measurements, the linewidth, FSS and radiative lifetime of exciton transitions are deduced.

\begin{figure}[!t]
\centering
\includegraphics[width=0.5\textwidth]{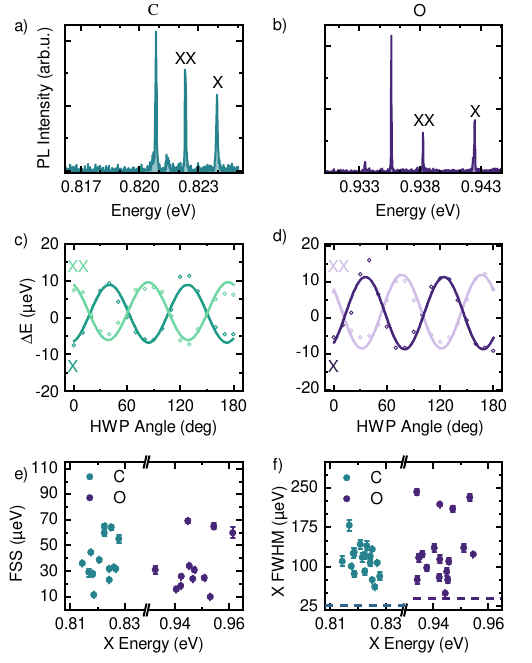}
\caption{a), b) $\mu$-PL spectra of sample \textbf{C} and sample \textbf{O}, recorded via above-bandgap excitation. c), d) Polarization-resolved $\mu$-PL measurements of samples \textbf{C} and \textbf{O}, respectively, as a function of the rotation angle of the half wave plate (HWP) under non-resonant excitation. The fine structure splitting (FSS) values for the QD emission of panels d) and c) are \SI{19.0}{ \micro eV} $\pm$ \SI{0.4}{ \micro eV} and \SI{14.0}{ \micro eV} $\pm$ \SI{0.5}{ \micro eV}, respectively. e) FSS values of X emissions as a function of the emission energy. f) Linewidths (FWHM) of measured neutral exciton (X) transitions on both samples as a function of the emission energy. The purple and green dashed lines indicate the resolution limits of the spectrometer at both wavelengths.}
\label{figure2}
\end{figure}

Figure \ref{figure2}a and \ref{figure2}b present representative $\mu$-PL spectra from sample \textbf{C} and sample \textbf{O}, respectively. Sharp emission lines resulting from different excitonic transitions are visible. Those peaks are attributed to neutral exciton (X), trion, and biexciton (XX) transitions. This assigment is made based on power- and polarization-dependent measurements. Importantly, as the DBR enhances the extraction efficiency it enables clean spectra with high signal-to-noise ratio (S/N), which allows for fitting the data with small errors.

Figure \ref{figure2}f shows the linewidths (full width at half maximum, FWHM) of X transitions from the emission spectra of $18$ individually characterized QDs for each telecom band. The extracted FWHMs values are in the range \SIrange{50.0}{243.0}{ \micro eV} with a median value of \SI{124.0}{ \micro eV} $\pm$ \SI{6.4}{ \micro eV} for sample \textbf{O} and \SI{119.0}{ \micro eV} $\pm$ \SI{6.8}{ \micro eV} for sample \textbf{C}. With the exception of some outliers, most data points are within the \SIrange{60.0}{150.0}{ \micro eV} range. The measured minima are \SI{50.0}{ \micro eV} $\pm$ \SI{0.9}{ \micro eV} for sample \textbf{O} and \SI{62.0}{ \micro eV} $\pm$ \SI{4.0}{ \micro eV} for sample \textbf{C}. This is close to the resolution limit of the spectrometer. For sample \textbf{O} (\SI{1310}{ nm}) the resolution limit is \SI{35.0}{ \micro eV}, while for sample \textbf{C} (at \SI{1550}{ nm}) it is \SI{23.0}{ \micro eV}. The resolution limits are illustrated as green and purple dashed lines, respectively.\\
Previous studies of MBE-grown SK QDs grown atop compositionally graded layers on (001) GaAs substrates report median values of FWHM of X emission of \SI{300.0}{ \micro eV} in the C-band \cite{wyborski2023impact} and of \SI{250.0}{ \micro eV} for a charged  exciton \cite{wronski2021metamorphic}.
For MBE-grown, DE QDs on (111)A GaAs substrates values in the range of \SIrange{100.0}{550.0}{ \micro eV} are reported \cite{barbiero2022exciton}.\\
Meanwhile, for MOVPE-grown QDs on InP substrates, Ref. \cite{vajner2024demand} reports values of \SI{119.0}{ \micro eV}, and Ref. \cite{holewa2024high} reports a mean value of \SI{557.2}{ \micro eV}.
Therefore, the FWHM values here reported are lower than those reported for other self-assembled, MBE-grown QDs on graded layers and on GaAs substrates of different orientations. We also observe an improvement over values measured on MOVPE-grown QDs on InP substrates. All the values mentioned above were obtained in measurements conducted under nonresonant excitation.\\
As the FSS values play a crucial role in realizing polarization entangled photon sources \cite{sapienza2013exciton, lettner2021strain, zeuner2021demand}, we carried out a polarization-resolved analysis of the emission of $13$ QDs under above-bandgap excitation for samples \textbf{O} and \textbf{C}. For each QD, spectra with a linear polarizer (LP) and half waveplate (HWP) in the detection channel were taken as a function of the angle of the half wave plates. The spectra were then fitted to extract the center position of the X and XX peaks. The results of two example measurements from samples \textbf{C} and \textbf{O} are presented in Figures \ref{figure2}c and \ref{figure2}d. The figures show the extracted deviation of the emission energy from its mean value as a function of the rotation angle of the HWP. In both cases, counter oscillating wavelengths for X and XX are observed. Sine fits (continuous lines) to the data reveal FSS values of \SI{19.0}{ \micro eV} $\pm$ \SI{0.4}{ \micro eV} (sample \textbf{O}, purple) and \SI{12.0}{ \micro eV} $\pm$ \SI{0.5}{ \micro eV} (sample \textbf{C}, green). 

The extracted values of the FSS for all investigated QDs are presented in
Figure \ref{figure2}e as a function of their emission energies. 
For the majority of the measured QDs we extract FSS values $\leq$ \SI{50.0}{ \micro eV}, with median values of \SI{28.0}{ \micro eV} $\pm$ \SI{1.6}{ \micro eV} for \textbf{O} and \SI{34.0}{ \micro eV} $\pm$ \SI{1.4}{ \micro eV} for \textbf{C}.
Importantly, the smallest observed values are \SI{10.0}{ \micro eV} $\pm$ \SI{0.4}{ \micro eV} for \textbf{O} and \SI{12}{ \micro eV} $\pm$ \SI{0.55}{ \micro eV} for \textbf{C}.

To asses our results, we compare the extracted values to those published in literature.
For QDs grown using the Stransky Krastanov (SK) approach, reported mean values are \SI{176.0}{ \micro eV} $\pm$ \SI{9.0}{ \micro eV} for MBE-grown QDs on InP \cite{skiba2017universal} and in the range \SI{123.0}{ \micro eV} for MOVPE-grown QDs on InP \cite{holewa2022bright}. For SK-grown QDs on GaAs substrates, where the FSS is affected by the higher lattice mismatch and consequent dot asymmetry, reported values for MBE-grown QDs range within \SIrange{16.0}{136.0}{ \micro eV}, which were reduced by applying external strain  \cite{sapienza2013exciton}, \SI{50.0}{ \micro eV} \cite{wyborski2023impact} for (001)-oriented substrates.
Meanwhile, for QDs grown using droplet epitaxy (DE), which results in lower FSS values due to the improved symmetry of the QDs, reported mean values are \SI{42.0}{ \micro eV} $\pm$ \SI{7.0}{ \micro eV} with a minimum of \SI{12.0}{ \micro eV} $\pm$ \SI{2.0}{ \micro eV} for MBE-grown QDs on InP substrates \cite{skiba2017universal} and \SI{50.0}{ \micro eV} $\pm$ \SI{5.0}{ \micro eV} for an MOVPE-grown QD on InP\cite{holewa2022droplet}. For MBE-grown QDs on (111)A-oriented substrates, values of \SI{141.2}{ \micro eV} $\pm$ \SI{4.2}{ \micro eV} are reported \cite{barbiero2022exciton}
Notably, the reported FSS values are lower than the ones reported for QDs grown on InP substrates via SK growth and remarkably comparable to DE-grown QDs on InP, and they are however comparable to other MBE- \cite{wyborski2023impact} and MOVPE- \cite{paul2017single} grown QDs on graded layers on top of (001) GaAs substrates. We tentatively attribute this to the lower strain generated by this growth approach, which should result in more symmetric QDs.

\begin{figure}
\centering
\includegraphics[width=0.5\textwidth]{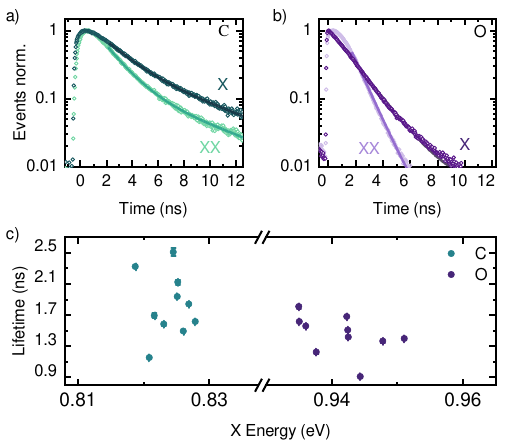}
\caption{a),b) Lifetime measurements of X and XX transitions on QDs from sample \textbf{C} and \textbf{O} recorded under non-resonant pulsed excitation.  
c) Lifetime of measured X transitions as a function of the X emission energy.}
\label{figure4}
\end{figure}

\section{Time-resolved measurements}
To fully characterize the optical properties of the QD emission, we quantify the excited state lifetime $\tau$ via pulsed non-resonant excitation.
Figures \ref{figure4}a and \ref{figure4}b present the time-resolved PL intensity of typical measurements of the X and XX transitions in samples \textbf{C} and \textbf{O}. They show exponential (Fig \ref{figure4}b) and biexponential (Fig \ref{figure4}a) decay behaviour. 
The extracted $\tau$ values exhibit the expected (\cite{bacher1999biexciton})  ratio of approximately $2$ between XX and X lifetimes. For instance, the measurements presented in Fig. \ref{figure4}b (sample \textbf{O}) exhibit a mono-exponential decay with $\tau^X = $\SI{1.70}{ ns} $\pm$ \SI{0.01}{ ns} and $\tau^{XX} = $\SI{0.80}{ ns} $\pm$ \SI{0.01}{ ns} . For sample \textbf{C} (Fig. \ref{figure4}a ), a biexponential decay function was required to accurately fit the experimental data for both transitions. This could be caused by the presence of additional processes related to e.g. defects and refilling of the QD by charge carrier trap states \cite{wyborski2023impact, nawrath2019coherence}.
The extracted values are $\tau^X = $\SI{2.31}{ ns} $\pm$ \SI{0.09}{ ns} and $\tau^{XX} = $\SI{1.63}{ ns} $\pm$ \SI{0.03}{ ns}. The subsequent slower decays amount to \SI{19.29}{ ns} $\pm$ \SI{1.38}{ ns} for X and to \SI{8.00}{ ns} $\pm$ \SI{1.50}{ ns} for XX.

Fig. \ref{figure4}c shows the X lifetime of $10$ QDs from each sample. The values lie within the range \SIrange{0.94}{2.50}{ ns} with median values of \SI{1.40}{ ns} $\pm$ \SI{0.01}{ ns} for \textbf{O} and \SI{1.80}{ ns}  $\pm$ \SI{0.02}{ ns} for \textbf{C}, respectively.
The extracted X lifetime minima are \SI{0.91}{ ns} $\pm$ \SI{0.02}{ ns} for \textbf{O} and \SI{1.15}{ns} $\pm$ \SI{0.02}{ ns} for \textbf{C}. Similar values are reported across the literature, without significant differences between SK- (\cite{wyborski2023impact, nawrath2019coherence, paul2017single, sittig2022thin}) and DE- (\cite{holewa2022droplet, phillips2024purcell}) grown QDs nor between GaAs- (\cite{paul2017single, nawrath2019coherence, sittig2022thin}) and InP- (\cite{holewa2022droplet, phillips2024purcell}) based structures. 
This indicates that the $\tau$ values we report do not significantly differ from those reported in other studies on non cavity-integrated QD, across different growth approaches, methods, and substrate compositions.

\section{Correlation measurements}
\begin{figure}[!t]
\centering
\includegraphics[width=0.5\textwidth]{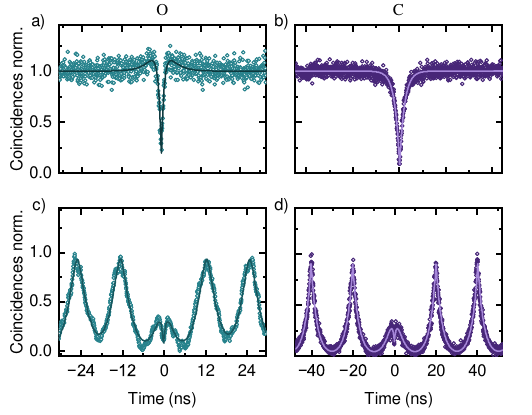} 
\caption{a),b) CW-excitation $\text{g}^{(2)}(0)$ measurements on samples \textbf{C} and \textbf{O} at excitation powers of \SI{30}{ nW} and \SI{5}{ \micro W} and. c),d) Pulsed-excitation  $\text{g}^{(2)}(0)$ measurements on \textbf{C} and \textbf{O}, carried out at a power of \SI{200}{ nW} and \SI{35}{ \micro W}. The difference in time scale between the two figures is due to exciting with two different pulsed lasers, with repetition rates of \SI{50}{ MHz} and \SI{80}{ MHz}, respectively.}
\label{figure5}
\end{figure}

Finally, we performed second-order autocorrelation measurements on trion states (selected due to their higher brightness over X states) to confirm the single-photon nature of the emission. The results of typical measurements for both samples under continuous-wave (CW) excitation are presented in Fig. \ref{figure5}a and \ref{figure5}b.
In both cases the measured $g^{(2)}(0)$ values are significantly below $0.50$, demonstrating the single-photon nature of the emitted light and a negligible background emission. By fitting we obtain $g^{(2)}(0)$ values of $g^{(2)}(0) = 0.08 \pm 0.01$ for sample \textbf{O} and $g^{(2)}(0) = 0.19 \pm 0.02$ for sample \textbf{C}.
Note that the data in Figure \ref{figure5}a exhibit a bunching behaviour, indicative of blinking. This blinking behaviour can be attributed, e.g., to trapping of charge carriers, re-excitation, or the presence of dark states \cite{nawrath2019coherence, holewa2022bright}.\\
\indent Typical second-order autocorrelation measurements under pulsed, above-band excitation are presented in Fig. \ref{figure5}c and \ref{figure5}d for sample \textbf{C} and \textbf{O}. Both measurements reveal an attenuated center peak with a sharp dip in its center. 
This behavior can be attributed to the non-resonant excitation scheme. A significant number of carriers are excited and trapped in the wetting layer or in other charge-trap states \cite{dalgarno2008hole, holewa2022droplet}. Upon a first recombination event, these carriers are subsequently captured by the QDs, resulting in the emission of secondary photons. Similar recapture dynamics have been observed in previous $g^{(2)} (0)$ measurements conducted under non-resonant excitation on QDs grown directly on compositionally graded InGaAs layers \cite{wyborski2023impact, veretennikov2025single, paul2017single}.

The presence of a visible dip in the center peak, combined with the results of the $g^{(2)} (0)$ measurements under CW excitation and the near absence of background, hint at this phenomenon being responsible for the counts near time zero of the $g^{(2)} (0)$ value under pulsed excitation. In particular, this would be due to the chosen excitation scheme and to the higher excitation power required to conduct measurements under pulsed excitation.
Recapture behavior as a function of quasi-resonant or resonant excitation will however be investigated in detail in future studies, as it is beyond the scope of this paper.
The different width and shape of the side peaks of Fig. \ref{figure5}c compared to Fig. \ref{figure5}d is caused by a rising onset in the population lifetime of the measured transition, likely originating from the near saturation excitation conditions and possible double excitation effects \cite{nawrath2019coherence}. On the other hand, the discrepancy in the spacing between them originates from the use of another excitation laser with a different repetition rate.\\
\indent These correlation measurements confirm the single-photon emission of the studied structure. Combined with the results from the linewidth and FSS measurement, they underline the potential of the studied sample structure.

\section{Conclusions and Outlook}
In this paper, we demonstrated the realization of high-quality single-photon emitters operating in the second and third telecom windows. Our design is based on InAs QDs embedded within a InGaAs matrix, grown on a compositionally graded InGaAs buffer to reduce dislocations. 
We found that the emission exhibits narrow linewidths even under non-resonant excitation and without implementing additional measures to control the charge environment, likely due to the improved dot symmetry and reduced dislocations in the QD matrix.
As further evidence of this growth approach's benefits in terms of strain and dot shape, the FSS values we measured are comparable to those reported for QDs grown with approaches and materials known for yielding symmetric QDs characterized by small FSS values, such as DE on InP substrates. 
Moreover, we confirmed the single-photon nature of the emission, although the excitation scheme employed in these measurement results in charge carrier recapture, which negatively affects our extracted $g^{(2)}(0)$ values.\\
\indent These results establish InGaAs-embedded quantum dots as strong candidates for single-photon sources in quantum communication systems operating within the telecom O- and C-bands. By minimizing the dislocation density in the matrix where the QDs are embedded, the presented growth approach will also open the pathway for the fabrication of electrically contacted InAs QDs embedded in GaAs-based resonators with telecom emission. \\
\indent To further advance towards quantum communication applications, a transition to resonant excitation schemes would be beneficial, since it would allow for better quantification of the transition linewidth and eliminate the limitations caused by recapture dynamics observed by the autocorrelation measurements.
Incorporating the heterostructure into piezoelectric actuators could further reduce the FSS, potentially enabling the use of the presented QDs as sources of entangled photon pairs.
Moreover, embedding the presented QDs into nanocavities to enhance generation rates will enable brighter and purer emission, thus paving the way for more advanced quantum communication protocols.

\section{Acknowledgements}
We gratefully acknowledge financial support from the Deutsche Forschungsgemeinschaft (DFG, German Research Foundation) via projects INST 95/1220-1 (MQCL) and INST 95/1654-1 (PQET), Germany’s Excellence Strategy (MCQST, EXC-2111, 390814868), the Bavarian Ministry of Economic Affairs (StMWi) via project 6GQT and the German Federal Ministry of Education and Research via the project 6G-life.

\section{Methods}
Ensemble PL spectroscopy of InAs QDs was performed under continuous wave non-resonant excitation at \SI{660}{ nm} by using a helium flow cryostat operating at \SI{10}{ K}. We analyzed the detected PL signal using a spectrometer equipped with a thermoelectrically (TE)-cooled InGaAs linear array detector. Reflectivity measurements were carried out by excitation with a tungsten halogen light source.  
The surface morphology was investigated by a Bruker Multimode 8 AFM in tapping mode.\\ 
\indent The optical characterization of single QDs was carried out using a conventional confocal microscopy $\mu$-PL setup. The samples were placed in an AttoDRY$800$ closed-cycle cryostat operating at \SI{4}{ K}. The excitation was provided by a non-resonant continuous-wave (CW) Toptica CTL laser at \SI{898}{ nm}. The PL signal was analyzed using an Andor Shamrock 750 spectrometer and detected by a an InGaAs CDD iDus 419, at a temperature of \SI{-66}{\celsius}. Polarization-resolved measurements were performed by detecting the emitted signal as a function of the rotation angle of a HWP positioned in front of a LP in the detection path.\\
\indent To enable time-resolved measurements, we used pulsed excitation from a Coherent Chameleon Ti:Sapphire laser, tuned to \SI{785}{ nm} with a repetition rate of \SI{80}{ MHz}. The laser pulses, originally in the femtosecond range, were stretched to the picosecond timescale using a custom-built 4f spectral filter. A NKT Photonics Origami pulsed laser, operating at the same wavelength with a repetition rate of \SI{50}{ MHz} was additionally used to perform second-order autocorrelation measurements.
The emitted transitions were spectrally filtered using a similar setup and detected by Single Quantum superconducting nanowire single-photon detectors (SNSPDs).
For photon correlation measurements, we added a Hanbury Brown and Twiss (HBT) setup, employing a fiber-coupled beam splitter from OZ Optics. 
Time-tagging of detected photons was implemented using a qutools QuTag Standard time-correlated single-photon counting (TCSPC) module.

\medskip

\end{document}